\documentclass[a4paper, oneside, 12pt, fleqn]{article}
\usepackage{arxiv}

\usepackage[utf8]{inputenc}
\usepackage{srcltx}
\usepackage{amsmath}
\usepackage{amssymb}
\usepackage{graphicx}
\usepackage{blindtext}
\usepackage[title]{appendix}
\usepackage[utf8]{inputenc}
\newcommand{\half}{\scriptstyle \frac{1}{2} \displaystyle}

\begin{document}
\pagestyle{myheadings}
\markright{Tube laws and wave speeds}

\title{Arterial Tube Laws and Wave Speeds}
\author{Kim H. Parker) \\
	Department of Bioengineering\\
	Imperial College \\
	London SW7 2AZ, UK \\
	\texttt{k.parker@imperial.ac.uk}}

\maketitle

\begin{abstract}
    The 1-D theory of flow in the arteries yields an equation for the wave speed in terms of the density of blood and the distensibility of the vessel. By means of this equation there is a duality between the equation for the wave speed and the tube law describing the area of the vessel as a function of pressure. We explore this duality for the equations for wave speed and tube laws that are most commonly used in theoretical arterial hemodynamics. We see that there are qualitative differences between these laws and the experimental data on wave speed in canine arteries measured by Anliker and his colleagues 50 years ago. We suggest an empirical equation for wave speed (and its dual tube law) that fits the experimental data as well as the common expectation that arteries become stiffer as the pressure increases. We conclude with a cautionary historical tale about the differences between the theoretical predictions and the experimental measurements of the speed of sound in air that persisted for more than 200 years.

\end{abstract}

\section{Introduction and background}

The local wave speed in an artery is directly related to the distensibility of the artery and the density of blood. This relationship is derived from the 1-D conservation equations (mass and momentum) using the method of characteristics and can properly be thought of as the definition of the wave speed. Historically there have been a number of equations for the wave speed in an artery derived for particular assumptions about the anatomy and mechanical properties of the arterial wall.

Experimentally there have been many studies of arterial wave speed. Most of these involve the measurement of the time between the arrival of the foot of the arterial pulse wave at different sites in the arterial system and the measured distance (or more usually the assumed distance) between the sites. Since these experiments measure some average of the wave speeds over the path of the wave they cannot be considered as measurement of a local wave speed.\footnote{Since wave speed is a ratio, the appropriate average over a path including different vessels with different wave speeds is the harmonic average (\textit{i.e.} the reciprocal of the average is equal to the average of the reciprocals.) This is seldom, if ever, done.}  Other methods have been suggested for calculating local wave speed from measurements of local flow and pressure or area or diameter. All of these methods are based on different assumptions and it is difficult to assess the validity of these methods in the absence of a 'gold-standard' method for measuring local arterial wave speed.

A singular exception to this statement about measurement of local wave wave speed in arteries is the work of Anliker and his colleagues during the late 1960s and early 1970s. These experiments, which will be described in more detail later in this paper, provide a body of evidence about wave propagation in dog arteries \textit{in vivo} that should probably be taken as 'gold-standard' experimental evidence.

This paper paper is based on a talk given at BioMedEng18 which was held at Imperial College, London, UK on 6-7 September 2018. It is not a thorough review of theory and experiments about arterial wave speed. Our intention is to review the most commonly used equations for arterial wave speed and the most commonly used arterial 'tube laws' and to compare them with experimental results, in particular the findings of Anliker \textit{et al}. We find serious discrepancies between theory and experiment which must be resolved before we can claim to understand wave speed in arteries. We will conclude with some suggestions for the use of empirical tube laws derived from the experimental evidence and explore what they reveal about the distensibility (i.e. the stiffness) of arteries.

\section{The definition of local wave speed}

The most rigorous derivation of the theoretical definition of wave speed comes from the method of characteristics solution of the 1-D equations for the conservation of mass and momentum in an artery. This is the basis of the following derivation. A very simple phenomenological derivation involving only simple algebra is given in the Appendix. 

Assuming that blood is incompressible, 1-D mass conservation requires
\[
 A_t + UA_x + AU_x = 0
\]
1-D conservation of momentum requires
\[
 U_t + UU_x = -\frac{1}{\rho}P_x + R
\]
where subscripts denote partial derivatives. The dependent variables are $A(x,t)$, the cross sectional area, $U(x,t)$, the mean velocity over the cross section, $P(x,t)$, the pressure and the independent variables are $t$, time and $x$, the axial distance. $R$ represents the net viscous stress acting on the fluid. These 2 equations involve 3 dependent variables and so it is necessary to assume a relationship between $A$ and $P$, generally referred to as the tube law. This can be written either in the form $A(P)$ or its inverse $P(A)$. For this derivation we will assume $A(P)$ and discuss the alternative form subsequently.

Assuming $A(P)$,  $A_t = A_PP_t$  and $A_x = A_PP_x$ where $A_P = \left(\frac{\partial A}{\partial P}\right)_{t,x}$. Substituting into the conservation equations and rearranging
\[
 P_t + UP_x + \frac{A}{A_P}U_x = 0
\]
\[
 U_t + \frac{1}{\rho}P_x + UU_x = R
\]
For the following derivation of wave speed, $R$ can depend on the dependent variables $P$, $A$ and $U$ and any other physical parameter such as viscosity but cannot be a function of the derivatives of any of the dependent variables. These equations can be written in canonical matrix form
\[
 \left(\begin{matrix} P \\ U \end{matrix}\right)_t + \left(\begin{matrix} U & \frac{A}{A_P}\\ \frac{1}{\rho} & U \end{matrix}\right) \left(\begin{matrix} P \\ U \end{matrix}\right)_x
    = \left(\begin{matrix} 0 \\ R \end{matrix}\right)
\]
The eigenvalues of the matrix are determined from the characteristic equation
\[
 \begin{vmatrix}U-\lambda & \frac{A}{A_P} \\ \frac{1}{\rho} & U-\lambda\end{vmatrix} = (U - \lambda)^2 - \frac{A}{\rho A_P} = 0
\]
Solving the characteristic equation the 2 eigenvalues are
\[
 \lambda = U \pm \left(\frac{A}{\rho A_P}\right)^{1/2} = (U \pm c)
\]
where we have defined
\begin{equation}
\label{eq:c_def}
 c^2 \equiv \left(\frac{A}{\rho}\frac{dP}{dA}\right)
\end{equation}
where $\frac{dP}{dA} = \frac{1}{A_P}$ is the total local derivative of pressure with respect to area. The solution of the equations by the method of characteristics follows from the observation that along the \textit{characteristic} directions $\frac{dx}{dt} = (U \pm c)$ the partial differential equations reduce to ordinary differential equations that can be solved using well-established methods.

The identification of the formally defined $c$ with wave speed follows from the observation that $c$ has the units of velocity. It is positive because $A$ and $\rho$ are positive physical parameters and $A_P \ge 0$ is a requirement for the mechanical stability of the artery (in a closed artery at pressure $P$, if $A_P < 0$ then any positive perturbation of $P$ would decrease the volume, increasing $P$, leading to the collapse of the artery). Finally, the ordinary differential equations in the characteristic directions dictate that any perturbation of $P$ and $U$ in a vessel with zero velocity will propagate at speed $c$ in the forward and backward direction. We therefore take (\ref{eq:c_def}) as the basic theoretical definition of the local wave speed.

The method of characteristics solution of the 1-D conservation equations also yields an expression for the relationship between the changes in pressure and velocity across a incremental wavefront, $\Delta P$ and $\Delta U$. This argument is fairly subtle and is based on the observation that the ODEs describing the solution along the characteristic directions $\frac{dx}{dt}=U \pm c$ can be written 
\[
 \frac{dR_\pm}{dt} = f(P,U,A) \qquad \mbox{where}\qquad \frac{dR_\pm}{dt} = \frac{dU}{dt} \pm \frac{1}{\rho c} \frac{dP}{dt}
\]
where $R_\pm$ are usually referred to as the Riemann variables. If we consider a single wave propagating in, say, the forward directions, $\frac{dx}{dt}=U+c$ with a difference $\Delta P$ and $\Delta U$ across the wave. When there are no backward waves these differences must be such that the backward Riemann variable satisfies $\frac{dR_-}{dt} = 0$. That is $\Delta P - \rho c \Delta U = 0$. As long as the waves are additive (the acoustic assumption), this argument can be used for any combination of waves, yielding the so called 'water hammer equations'
\begin{equation}
\label{eq:water_hammer}
 \Delta P_\pm = \pm \rho c \Delta U_\pm
\end{equation}
where $\pm$ indicates the direction of travel of the waves.

It is important to note that $c$ can be a function of pressure $c=c(P)$ and that this function will depend on the tube law $A = A(P)$. The water hammer equations are still valid as they only depend on the additivity of the waves, but are more difficult to apply when $c$ is not constant.

\section{Theoretical tube laws and wave speeds}

The theoretical definition of wave speed given in (\ref{eq:c_def}) indicates the relationship between the wave speed and the tube law governing the area of the vessel as a function of pressure. If we have an equation for the wave speed as a function of pressure $c = c(P)$ we can substitute it into (\ref{eq:c_def}) and separating variables, write
\[
 \frac{dA}{A} = \frac{dP}{\rho c(P)^2}
\]
retaining the form $c(p)$ as a reminder that $c = f(P)$. Integrating both sides of the equation
\begin{equation}
\label{eq:lnA}
 \frac{A}{A^*} = e^{\int_{P^*}^P \frac{dP}{\rho c(P)^2}}
\end{equation}
where $P^*$ is any convenient reference pressure and $A^* = A(P^*)$. 

Conversely, given any tube law $A=A(P)$ we can calculate $\frac{dA}{dP}$ and substitute into (\ref{eq:c_def}) to find the wave speed equation $c = c(P)$. Therefore, we should treat the tube law and the equation for the wave speed as dual relationships.

Although arteries have a complex, multi-component structure, most theoretical tube laws are derived assuming the arterial wall is homogeneous and elastic. Curiously, there are two different results obtained by slightly different approaches to this basic problem in solid mechanics which are currently in use; a solution for tubes with arbitrary dimensions and and a solution for thin-walled tubes using Laplace's law. We will briefly describe both approaches and compare their results when applied to the calculation of the wave speed given in (1).

\subsection{thick-walled vessels}

We start the discussion of tube laws with the general case of an infinite, axisymmetric, uniform, homogeneous elastic vessel. In the stress-free state the vessel has inner radius $a$, outer radius $b$ and the wall properties are assumed to be linearly elastic with modulus E and Poisson's ratio $\nu$. We note that the assumption of an infinite uniform tube is equivalent to assuming that the axial length is constant. The equation for the radial equilibrium displacement of the wall in response to a luminal pressure $P_a$ and and external pressure $P_b$ is \cite{Bower2009}
\[
 \frac{\partial}{\partial r}\left(\frac{1}{r}\frac{\partial}{\partial r}(ru)\right) = 0
\]
where $u$ is the radial displacement and $r$ is the radial coordinate. The general solution is
\[
 u(r) = \Theta r +\frac{\Psi}{r} \qquad \mbox{with} \qquad \Theta = \frac{(1+\nu)(1-2\nu)}{E}\frac{P_aa^2 - P_bb^2}{b^2-a^2}, \quad
 \Psi = \frac{(1+\nu)}{E}\frac{a^2b^2(P_a-P_b)}{b^2-a^2}
\]
If we further assume that the wall is incompressible, $\nu = 1/2$ and $\Theta =  0$ giving the radial displacement in terms of the transmural pressure $P = P_a-P_b$
\[
 u(r) = \frac{3a^2b^2}{2E(b^2-a^2)}\frac{P}{r}
\]
Defining $A$ as the area of the lumen, $A = \pi(a+u(a))^2$ and the unstressed area of the lumen is $A_0 = \pi a^2$. Substituting and expressing $P$ as a function of $A$
\[
 P = \frac{2E}{3}\frac{b^2-a^2}{b^2}\left(\left(\frac{A}{A_0}\right)^{1/2} -1\right)
\]
For comparison with the thin-walled case, it is convenient to define the wall thickness in the unstressed configuration $h=b-a$ and to write this equation in terms of the wall thickness ratio $\frac{h}{a}$. Observing that
\[
 \frac{b^2-a^2}{b^2} = \frac{h}{a}\left(\frac{1+\frac{1}{2}\frac{h}{a}}{1+2\frac{h}{a}+\left(\frac{h}{a}\right)^2}\right) 
\]
and recalling that $A_0 = \pi a^2$ we can write
\[
 P = \frac{4\sqrt{\pi}Eh\phi}{3\sqrt{A_0}}
 \left(\sqrt{\frac{A}{A_0}}-1\right) \qquad \mbox{where} \qquad \phi \equiv \left(\frac{1+\frac{1}{2}\frac{h}{a}}{1+2\frac{h}{a}+\left(\frac{h}{a}\right)^2}\right) 
\]
Finally, this can be written conveniently as
\[
 P = \frac{\beta \phi}{A_0} \left(\sqrt{A}-\sqrt{A_0}\right) \qquad \mbox{where} \qquad \beta \equiv \frac{4\sqrt{\pi}Eh}{3}
\]

\subsubsection{the $\beta$-law for thin-walled vessels}

The above formulation of the thick-walled tube law is convenient because we can see that for $\frac{h}{a} \ll 1$, $\phi \rightarrow 1$ and we obtain the thin-walled result to $\mathcal{O}(\frac{h}{a})$
\[
 P = \frac{\beta}{A_0}\left(\sqrt{A}-\sqrt{A_0}\right) 
\]
This tube law has been used extensively in 1-D computer modelling of the human arterial systemand is generally referred to as the $\beta$-law. (For an extensive review of tube laws used in computational arterial mechanics see \cite{Maynard2008}).

Because arteries in normal physiological conditions never experience stress-free conditions, it is convenient to use a more physiological state as the reference condition for the tube law. An obvious candidate would be diastolic pressure, but we will consider an arbitrary  reference pressure $P^*$ and define the reference area $A^* = A(P^*)$. We also find it convenient to define a reference wave speed $c^* = c(P^*)$. Applying this idea to the thick-walled tube law (and by setting $\phi =1$, to the thin-walled tube law) we can write
\[
 P - P^* = \frac{\beta \phi}{A_0} \left(\sqrt{A} - \sqrt{A^*}\right)
\]

Using the theoretical definition of the wave speed (\ref{eq:c_def})
\[
 c^2 = \frac{AdP}{\rho dA} = \frac{\beta \phi \sqrt{A}}{2\rho A_0} \qquad \mbox{and} \qquad
 c^{*2} = \frac{\beta \phi \sqrt{A^*}}{2\rho A_0}
\]
In terms of the reference wave speed the tube law and the equation for the wave speed squared is
\[
 c^2 = c^{*2}\sqrt{\frac{A}{A^*}}
\]
and the tube law is
\[
\frac{A}{A^*} = \left(1 + \frac{P - P^*}{2\rho c^{*2}}\right)^2
\]
This form of the equations will be convenient for comparing results for different theoretical expression and, particularly, for comparisons with experimental data.

\subsection{thin-walled tubes using the law of Laplace}

An alternative tube law has been derived for thin-walled, homogeneous elastic vessels using the law of Laplace as the starting point of the analysis. Curiously, this derivation gives results that are different from the $\beta$-law derived in the previous section although all of the basic assumptions about the geometry and material properties of the vessel wall are the same. The derivation presented here follows \cite{Olufsen2000}. Laplace's law balances the circumferential stress, $\tau$, in the walls and the transmural pressure stress
\[
 \tau = \frac{rP}{h} = \frac{E}{(1 - \nu^2)}\frac{(r - a)}{a}
\]
where $a$ is the unstressed radius of the vessel and $\frac{r-a}{a}$ is the circumferential strain, E is the elastic modulus, and $\nu$ is the Poisson ratio. For an incompressible wall $\nu = 1/2$. Solving for the transmural pressure
\[
 P = \frac{4\sqrt{\pi}Eh}{3}\left(\frac{1}{\sqrt{A_0}} - \frac{1}{\sqrt{A}}\right)
\]
where we have again used $A_0 = \pi r_0^2$. As in the discussion of the thick-walled vessel it is convenient to chose reference conditions other than the stress-free configuration, $P^*$, $A^* =A(P^*)$ and $c^* = c(P^*)$. In terms of these arbitrary reference conditions
\[
 P - P^* = \beta \left(\frac{1}{\sqrt{A^*}}-\frac{1}{\sqrt{A}}\right)
\]
where we have used the previous definition $\beta = \frac{4\sqrt{\pi}Eh}{3}$. 

As pointed out by Olufsen \textit{et al.} this result predicts that $\frac{dP}{dA}$ decreases with increasing $P$ (or $A$), contrary to the behaviour of real arteries in which $E$ is not constant but increases as the arterial wall is stretched.\cite{Olufsen2000} Despite this it is instructive to look at the equation for the wave speed resulting from this tube law. From the theoretical definition of wave speed
\[
 c^2 = \frac{AdP}{\rho dA} = \frac{\beta}{2\rho \sqrt{A}} \qquad \mbox{and} \qquad c^{*2} = \frac{\beta}{2\rho \sqrt{A^*}}
\]
In terms of $c^*$
\[
 c^2 = c^{*2} \sqrt{\frac{A^*}{A}}
\]
and
\[
\frac{A}{A^*} = \left(1 - \frac{P - P^*}{2\rho c^{*2}}\right)^{-2}
\]

\subsubsection{Moens-Korteweg equation}

An informal survey of medical and physiologal texts suggests that the most commonly used equation for the arterial wave speed is the Moens-Korteweg equation proposed independently by Moens and Korteweg in 1878\footnote{This formula was originally proposed by Thomas Young in 1808 using a semi-rigorous derivation based on Newton's derivation of the speed of sound in air.\cite{Young1808}}
\[
 c^2 = \frac{Eh}{2\rho r}
\]
A variant of this equation for a tethered artery (i.e. constrained so there is no axial movement of the wall) is also commonly used \cite{Fung1981}
\[
 c^2 = \frac{Eh}{2(1-\nu^2)\rho r}
\]
where $\nu$ is the Poisson ratio of the vessel wall. For an incompressible wall $\nu = 1/2$ and the tethered Moens-Korteweg equation becomes
\[
 c^2 = \frac{\beta}{2\rho \sqrt{A}}
\]
using $\beta$ as defined previously. This is the equation for the wave speed derived for a thin-walled vessel using Laplace's law. Using the duality of the equation for wave speed and the tube law, we conclude that the Moens-Korteweg equation implies the thin-walled tube law derived using Laplace's equation.

\subsubsection{wave speed constant}

Although it is not usually thought of as an equation for the wave speed, the most common assumption about arterial wave speed is that it is constant (i.e. not a function of pressure). This is frequently assumed in theoretical models of the circulation, usually without comment. If $c^*$ is constant we can substitute this in (\ref{eq:lnA}) to give find the corresponding tube law 
\[
 \frac{A}{A^*} = e^{\frac{P-P^*}{\rho c^{*2}}}
\]

\subsection{summary of commonly used equations for wave speed and tube laws}

We conclude this section with a table of the most commonly used equations for arterial wave speed and their related tube laws, remembering that the two are related through the definition of wave speed (\ref{eq:c_def}). We write these equations normalised by the reference wave speed  $c^* = c(P^*)$ and the reference area $A^* = A(P^*)$ where $P^*$ is any convenient reference pressure.

The qualitative difference in the tube laws for thin-walled vessels (the thin-walled $\beta$-law and the Moens-Korteweg equation) is striking: the $\beta$-law predicts that wave speed increases with increasing pressure and the tube law associated with the Moens-Korteweg equation predicts that it decreases with increasing pressure. The reason for this difference is not immediately clear since both are derived under exactly the same assumptions; an axisymmetric, homogeneous, elastic and thin wall. I may lie in the way that the thin-wall approximation is implemented in the two analyses. In the derivation of Laplace's law it is assumed that the displacement of the wall is small relative to the unstressed radius of the vessel, $\frac{u(a)}{a} \ll 1$. In the analysis leading to the $\beta$-law there is no limitation on the displacement of the wall, only that the unstressed wall thickness is small relative to the unstressed radius, $\frac{h}{a} \ll 1$. Since the circumferential wall stress diverges as $h \rightarrow 0$, the first approach places limits on the ratio of $\frac{P}{E}$ which are not required under the second approach. This suggests that the $\beta$-law formulation should be preferred and that the validity of the Moens-Korteweg equation for wave speed as a function of pressure may be limited.

The commonly used equations for the wave speed (the first three entries in the table) can be expressed in terms of a single parameter calculated from the properties of the vessel. This parameter can be absorbed into the reference wave speed $c^*$ and so the equations and tube laws can be written in terms of only the density of blood $\rho$ and the reference variables $P^*$, $A^* = A(P^*)$ and $c^* = c(P^*)$. In terms of these parameters the thick and thin-walled homogeneous elastic cases have identical functional forms. The empirical exponential law has an independent parameter $a$ which means there is a family of wave speed equations and tube laws with different values of this parameter.

\begin{center}
\begin{table}[ht]
\centering
\begin{tabular}{|c|c|c|c|} \hline
 {\rule[-3mm]{0mm}{10mm}}tube law & assumptions & $\frac{c(P)}{c^*}$ & $\frac{A(P)}{A^*}$ \\ \hline
 {\rule[-4mm]{0mm}{12mm}}$\beta$-law & thick or thin wall & $\left(1 + \frac{P-P^*}{2\rho c^{*2}}\right)^{1/2}$ & $\left(1 + \frac{P-P^*}{2\rho c^{*2}}\right)^2$ \\ \hline
 {\rule[-4mm]{0mm}{12mm}}Moens-Korteweg & thin wall Laplace & $\left(1 - \frac{P-P^*}{2\rho c^{*2}}\right)^{1/2}$ & $\left(1 - \frac{P-P^*}{2\rho c^{*2}}\right)^{-2}$ \\ \hline
 {\rule[-2mm]{0mm}{9mm}}constant  & - - - & 1 & $e^{\left(\frac{P-P^*}{\rho c^{*2}}\right)}$ \\ \hline \hline
 {\rule[-4mm]{0mm}{12mm}}exponential & empirical & $e^{a (P - P^*)}$ & $e^{\frac{1-e^{-a(P-P^*)}}{2a\rho c^{*2}}}$ \\ \hline 
\end{tabular}
\caption{\footnotesize{The most commonly used wave speed equations and their associated tube laws. The last entry is the empirical exponential law proposed in the final section of the paper. $a$ is an empirical constant with dimensions $Pa^{-1}$}}
\end{table}
\end{center}

\section{Experimental measurements of local arterial wave speed}

Clinical measurements of wave speed generally measure the time delay between the feet of cardiac pulse waves measured at two different sites. Generally these sites are chosen to be as far from each other as possible (e.g. the common carotid and femoral arteries) to maximise the time delay. This time and an estimate of the distance between the sites provides a measure of the pulse wave velocity. This measure is, of course, not a local wave speed but a weighted average wave speed in all of the arteries in the path (the wave spends more time in the arteries with lower waves speeds and this affects the form of the average). This measure has been shown to be a risk factor for various cardiovascular outcomes and is becoming an increasingly valuable clinical tool.\cite{Ben-Shlomo2014}

Local wave speed depend on the local properties of the artery and thus could be useful clinically although I do not know of any clinical study that has demonstrated this. Knowing the local wave speed, however, does allow us to apply further analysis to measurements of $P$ and $U$ to ascertain the forward and backward components of the arterial wave forms.\cite{Hughes2009} This information has proven to be very useful clinically and it is clear that it is important to find robust, accurate ways to measure local wave speed in arteries.

Local wave speeds are very difficult to measure clinically. Because wave speeds are relatively high in arteries (on the order of 10 m/s) and the lengths of individual arteries is relatively low (generally less than 10 cm) it is difficult to make direct measurements of the time delay of the cardiac pulse wave. As a result virtually all reported local wave speeds are based on a theoretical model involving the simultaneous measurement of $P$ and $U$ at a single site. The models that are used to derive the local wave speed are, in general, the subject of this study and so we cannot use these data to verify the models that we are discussing.

\begin{figure}[h]
\centering
 \includegraphics[width=12cm]{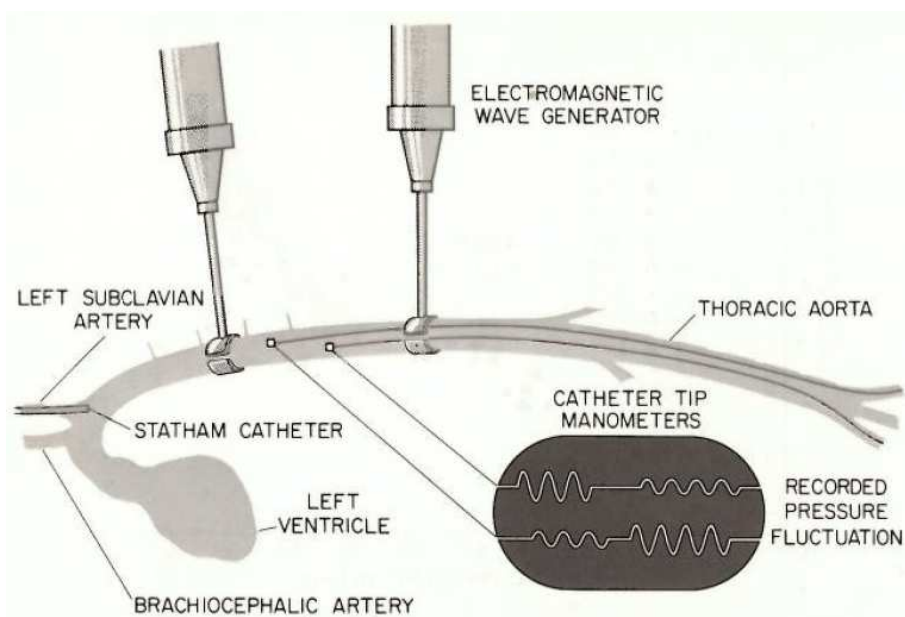}
\caption{\footnotesize{Schematic of the experimental measurement of wave speed in the descending aorta of a dog by Anliker and his colleagues. Short duration sinusoidal oscillations at relatively high frequencies are generated by either the upstream or downstream wave generator. The waves are detected by two catheter tip pressure transducers separated by a known distance. The amplitude and phase of the generated wave can be determined accurately using Fourier analysis. The phase determines the wave speed and the amplitude determines the dissipation. The waves can be generated at any time in the cardiac cycle relative to the R-wave in the ECG.\cite{Anliker1968}}}
\label{fig:Anliker_schematic}
\end{figure}

Anlicker and his colleagues generated a large volume of experimental \textit{in vivo} data about waves in the arteries of dogs.\cite{Anliker1968}\cite{Anliker1972}\cite{Histand1973} The experiments were ingenious, highly informative and thorough; anyone interested in arterial mechanics should be be fully aware of these exemplary studies. They looked at almost every facet of wave motion in arteries; types of waves, dispersion and dissipation of waves, whether the waves are convected by the fluid velocity in the artery and, of primary interest to this work, the effect of pressure on wave speed.

They showed that the dominant wave in dog arteries is the axisymmetric, axial waves that are discussed here. They used a number of ingenious methods in their studies and looked a the waves in a number of different arteries. We will look in particular at their study of waves in the descending aorta of the dog.\cite{Histand1973} The experimental setup that they used is shown in Figure \ref{fig:Anliker_schematic} and consists of two catheter tip pressure transducers advanced into the descending aorta from the two femoral arteries and positioned a known distance apart. Two electromagnetic wave generators were attached around the artery upstream and downstream from the pressure transducers. These wave generators caused transverse oscillation of the arterial walls with a controllable frequency and duration and could be triggered at any time during the cardiac cycle, as measured by a simultaneous ECG. These vibrations generated axial wave trains at the forcing frequency which propagated upstream and downstream in the artery and could be detected by the pressure transducers in the lumen. As shown in the inset of the figure, these short wave trains appeared in the output of the pressure transducers with a phase difference that depended on whether they were produced by the upstream or downstream wave generator. This phase difference could be determined very accurately by Fourier analysis of the transducer output. This phase difference and the known distance between the transducers allowed them to calculate the apparent wave speed (wave speed plus convective velocity). The direction of the wave could also be determined from the pressure traces because it was observed that these added waves dissipated with the amplitude decreasing in the direction of travel of the wave.\footnote{Annliker \textit{et al.} report many results about the dissipation of their waves showing that the amplitude of the waves decreased exponentially with distance measured in wavelengths of the generated wave trains $a = a_0e^{kx/\lambda}$. The dissipation constant $k$ is reported in several papers for forward travelling waves (in the direction of mean blood flow). In one paper they state that the attenuation (\textit{i.e.} dissipation constant, in the backward direction was always greater than in the forward direction, by a factor of 1.5-2.\cite{Histand1973} They suggest that this may be due to taper of the vessel, but present no evidence for this suggestions. I believe that a proper theoretical study of this observation could make a major contribution to our understanding of wave mechanics in arteries.}

\begin{figure}[ht]
\centering
 \includegraphics[width=12cm]{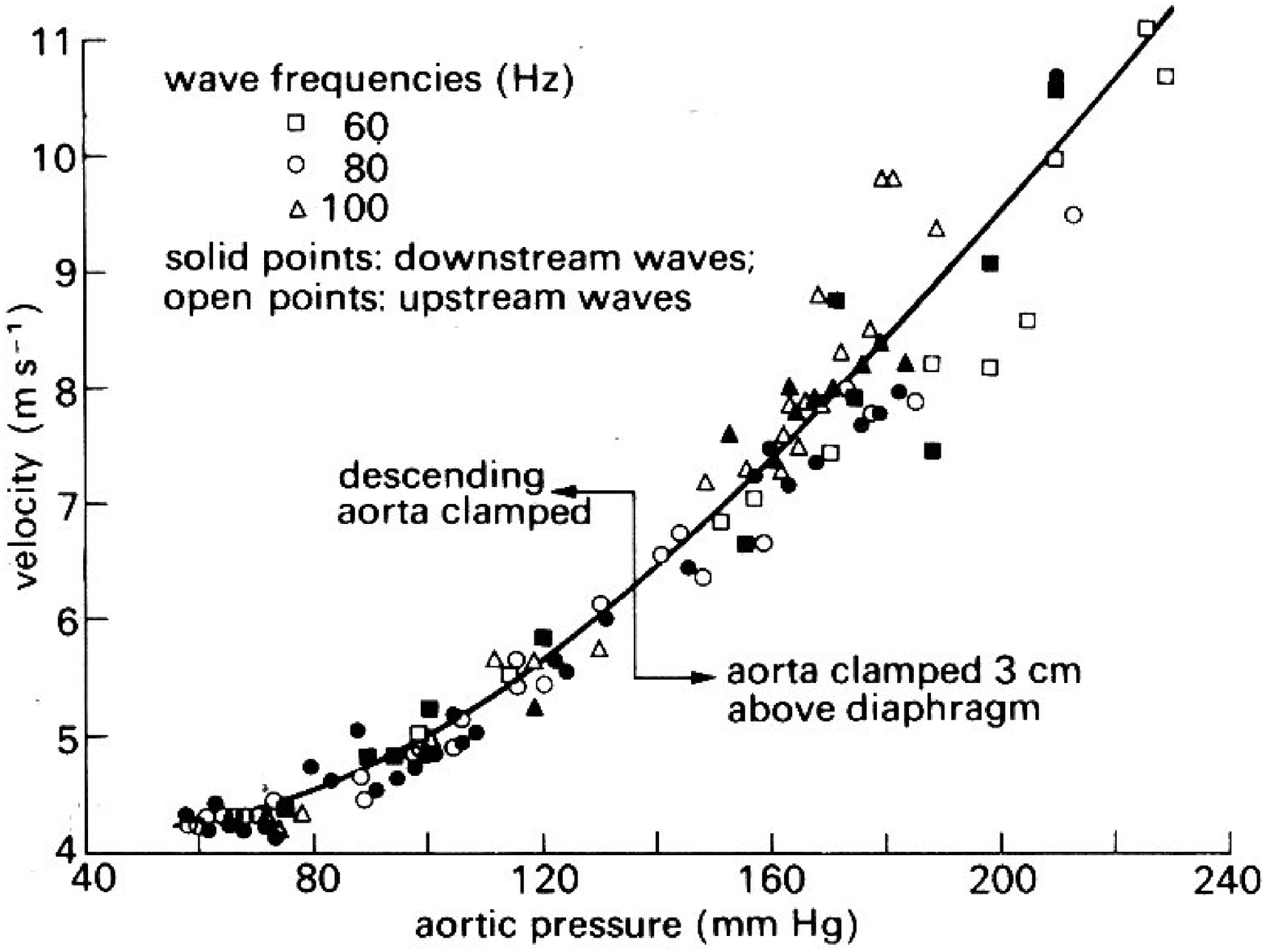}
\caption{\footnotesize{Experimental wave speed measured in the descending aorta of a dog as a function of pressure.\cite{Histand1973}}}
\label{fig:Anliker_data}
\end{figure}

In separate studies they demonstrated that the axial waves are convected by the mean velocity in the artery. By comparing the apparent wave speeds in the upstream and downstream direction they were able to eliminate the convective velocity to find the true wave speed. Because the wave trains were short, this experiment could be repeated at different times during the cardiac cycle and so the effect of pressure could be studied. To extend the range of pressures they made the dogs hypotensive by clamping the descending aorta upstream of the site of measurement and hypertensive by clamping the aorta above the diaphragm causing the renal control mechanism to cause hypertension.

The results of many repeated experiments in the descending aorta of a single dog are shown in Figure \ref{fig:Anliker_data}. The figure gives an indication of the care with which the experiments were performed. The solid points represent measurements for forward travelling waves and the open points backward waves. The squares, circles and triangles indicate the frequency of the generated waves. This figure shows data from approximately 100 separate measurements and they show very clearly that the wave speed is a monotonically increasing function of the pressure. Over the physiological range of arterial pressure (80 to 120 mmHg) we see that the wave speed increases from approximately 4.5 to 5.5 m/s. Over the whole range of pressure studied (60 to 220 mmHg) the wave speed increase from approximately 4.3 to nearly 11 m/s. These results are consistent with other studies of the effect of pressure on wave speed and, in general, this pattern is similar in different arteries, different individuals and in different species. We will take these data as the basis for our discussion in the rest of this paper.

\section{Comparison of experimental and theoretical wave speeds}

\begin{figure}[ht]
\centering
 \includegraphics[width=12cm]{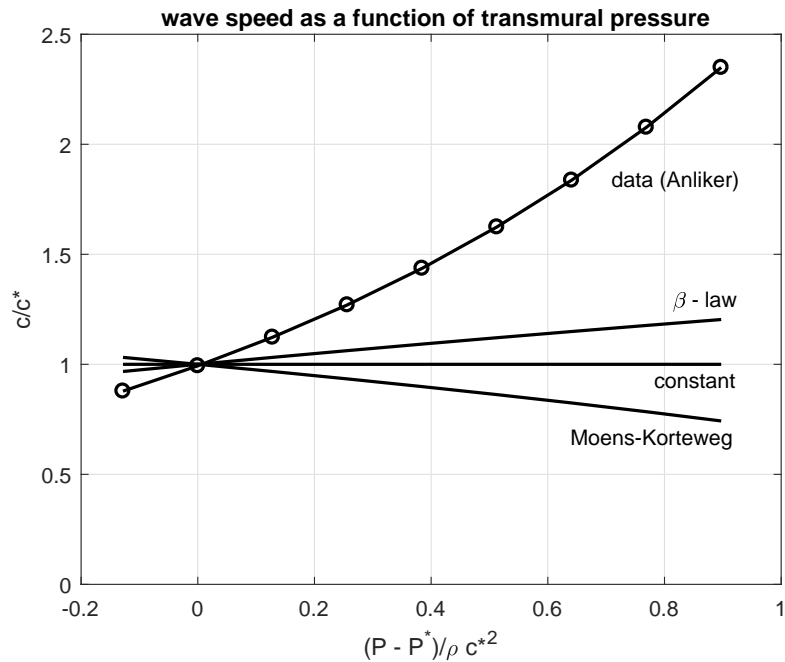}
\caption{\footnotesize{The normalised wave speed $c/c^*$ as a function of the normalised transmural pressure $P_t/\rho c^{*2}$. The black line with circular symbols is a fit to the experimental data shown in Figure \ref{fig:Anliker_data}. The solid curves represent the $\beta$-law, constant wave speed and the tube law corresponding to the Moens-Korteweg equation as labelled.}}
\label{fig:data_theory}
\end{figure}

Using the commonly used tube laws and equations for wave speed expressed in terms of the reference $P^*$, $A^*$ and $c^*$, we can compare them with the experimental results of Anliker \textit{et al.}\cite{Histand1973}. The results are shown in Figure \ref{fig:data_theory} where we have taken $P^* = 80$ mmHg and $c^* = 4.56$ m/s. It is clear from the figure that none of the commonly used tube laws are faithful to the experimental results; The $\beta$-law does predict an increase in wave speed with pressure, but it is concave to the pressure axis unlike the experimental results which are convex, the assumption of a constant wave speed does not fit the monotonically increasing experimental wave speed and the Moens-Korteweg law predicts that the wave speed will decrease with increasing pressure. These results provide a strong argument for the need for a better tube law to describe the mechanical properties of the arteries.

There have been a number of studies of the stress-strain behaviour of arterial wall tissue. Some are based on either uni- or bi-axial measurements of stress and strain and others based on the strain energy function governing arterial wall tissue.\cite{Fung1981} 

A very recent paper has addressed the limitations of the Moens-Korteweg equation for wave speed and suggested improvements to the analysis.\cite{Ma2018} They derive an expression for $P$ as as function of $A$ for the case of a linear elastic wall expressed in terms of the transcendental dilogarithm function, which makes it difficult to apply the results practically. Interestingly, this improved analysis also predicts that the $c$ will decrease with increasing $P$. They also solve the problem using the Fung hyperelastic model \cite{Fung1981} and obtain an expression for $P$ as a function of $A$ in terms of the imaginary error function. In this case they show that $c$ increases with increasing $P$ but, contrary to experimental observations, the rate of increase decreases with increasing $P$.

This and earlier studies are worthy of attention although all of them presume that the distensibility of the artery is determined solely by the arterial wall. \textit{In vivo} the distensibility of many arteries will also be affected by the interaction of the artery with its surrounding tissue. All arteries in the body are adjacent to other tissue (muscle, bone, veins, etc.) which will affect their distensibility. Obviously these interactions are extremely difficult to analyse theoretically. For this reason, we will look at feasible empirical tube laws and and their implications for arterial wave speeds.

\section{A suggestions for an empirical tube law and wave speed equation}

An obvious way to incorporate experimental measurements of wave speed as a function of pressure would be to fit the data to a polynomial using least squares. This, however, can lead to analytical problems in the derivation of the tube law arising from the intregrals of inverse polynomial functions. An analytically more convenient approach is to fit the experimental data with an exponential function. We explore this approach using the wave speed measurements of Annilker and his colleagues shown in Figure \ref{fig:Anliker_data}.

Assume that the wave speed is given by
\[
 c = c^* e^{a(P-P^*)}
\]
where $P$ is the transmural pressure and $c^*$ is the wave speed at $P=P^*$. Substituting into (\ref{eq:lnA}) we obtain the integral relationship
\[
 \frac{A}{A^*} = e^{\int_{P^*}^P \frac{e^{-2a(P-P^*)} dP}{\rho c^{*2}}} = e^{\frac{1 - e^{2a(P - P^*)}}{2a\rho c^{*2}}}
\]
Defining the dimensionless parameter $\alpha = 2a\rho c^{*2}$ we obtain
\[
 \frac{A}{A^*} = e^{\frac{1 - e^{-\alpha (p-p^*)}}{\alpha}}
\]
\begin{figure}[ht]
\centering
 \includegraphics[width=12cm]{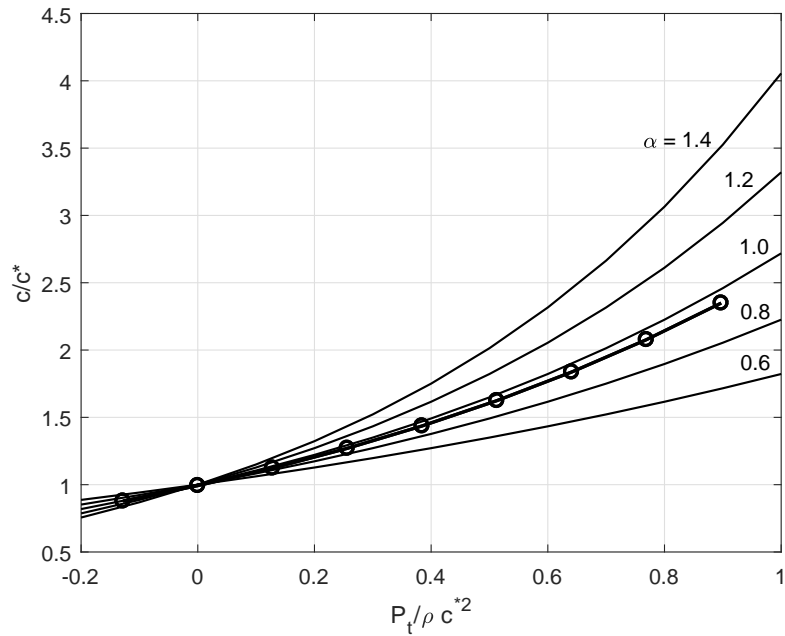}
\caption{\footnotesize{The normalised wave speed $c/c^*$ as a function of the normalised transmural pressure $P/\rho c^{*2}$. The black line with circles is a fit to the experimental data shown in Figure \ref{fig:Anliker_data}. The solid curves correspond to the indicated values of $\alpha$.}}
\label{fig:exp_law}
\end{figure}
These results are shown graphically in Figure \ref{fig:exp_law} for a representative range of the parameter $\alpha$. We see that this empirical form behaves in the expected way; the wave speed is convex to the pressure axis and the area ratio is concave to the pressure axis indicating that the vessel is getting stiffer as it dilates due to an increase in the transmural pressure. We also see that $\alpha = 1$ fits the experimental data of Anliker with good accuracy. A better estimate of $\alpha$ could easily be obtained by interpolation, but this does not affect the argument that an empirical exponential law for $c(P)$ gives satisfactory results.

In addition to fitting the Anliker data, there is further experimental evidence for an exponential relationship between wave speed and arterial pressure in \cite{Hughes1979}. Their results require some analysis because they report that the elastic modulus $E = E_0e^{aP}$ based on their interpretation of their ultrasound measurements of wave speed in the descending thoracic aorta and ascending aorta in 12 dogs. Unfortunately they do not give their measured wave speed data, only the elastic modulus calculated from the Moens-Korteweg equation. Working backwards through their method it is clear that their measured wave speed could be fitted with an exponential dependence on the pressure although it is impossible to derive trustworthy quantitative results because of an error in the Moens-Korteweg equation cited in the paper.\footnote{The Moens-Kurteweg equation (their equation 6) is given as $V_p = \frac{Et}{\rho d}$ where $V_p$ is the wave speed. The equation can be corrected by squaring $V_p$ but it is impossible to tell if this is simply a typographical error or if the wrong equation was used in the data analysis.} In any case, their measured wave speeds must have have depended exponentially on the pressure.

 \section{Conclusions and Discussion}

Arterial mechanics has benefited from the prevalence of cardiovascular disease in humans; it is generally accepted to be the major single cause of death world-wide.\cite{WHO2018} The role of mechanics and hemodynamics in the initiation and development of arterial disease has been the subject of intense study for more than half a century. Much has been learned but the clinical results of this understanding has been disappointing; almost all of the improvements in cardiovascular health have been the result of changes in public health and public awareness of various risk factors such as smoking and obesity. In order to make further progress in the prevention and treatment of cardiovascular disease it seems clear that we need better understanding of arterial mechanics and, perhaps, a re-evaluation of what we know and what we think we know. This is particularly true in the study of waves in the arteries.

 The 1-D theory of flow in elastic vessels indicates that there is a one-to-one relationship between the tube law describing the cross-sectional area of the vessel as a function of pressure and the equation describing the wave speed as a function of pressure. This is not a new observation but it is one that has been overlooked regularly. For example, almost all methods that have been described for the experimental determination of the local wave speed clinically make the implicit assumption that $c \ne f(P)$.
 
 Almost everyone working in arterial hemodynamics 'knows' that the local wave speed is a function of the location and type of artery; arteries are frequently classified as 'elastic' or 'muscular' and the distribution of elastin and collagen in different arteries has been studied intensively. Similarly, almost everyone knows that 'stiffening' of the arteries is caused by atherosclerosis and that this is a 'bad' thing. Unfortunately, the scientific basis of this knowledge is questionable. From the theoretical definition of wave speed (\ref{eq:c_def}) it follows that decreasing stiffness does increase the local wave speed, if we identify stiffness as distensibility $\frac{1}{a}\frac{dA}{dP}$. The global effects of local increases in wave speed are sadly lacking; even their collective effect on the distribution of pressure throughout the arterial system is largely unknown.
 
 Clinical studies of the effects of hypertension are informative but inconclusive. Currently they provide us with a wealth of correlations (and motivations) but little information about causal relationships. One of the most striking things about  hypertension is how deleterious even a relatively small increase in arterial pressure can be, particularly in the context of the large range of pressures the healthy cardiovascular system generates during the cardiac cycle and during normal activities such as exercising and sleeping. Why a relatively small increase in this normally widely fluctuating parameter causes long term derangement is a pressing but unanswered question.
 
 Researchers interested in arterial mechanics have made many advances during the past half century, but there is need for a deeper understanding of the subject if we are to contribute to the continuing epidemic of cardiovascular disease. As we have demonstrated, the most commonly used equations for the local wave speed, a critical parameter in arterial mechanics, are deficient. The Moens-Korteweg equation may give a reasonable value at a given pressure but it implicitly predicts that wave speed will decrease with increasing arterial pressure, which is generally observed to be wrong. The equation for wave speed derived from the commonly used $\beta$-law describing the area of a vessel as a function of pressure does predict that the wave speed will increase with increasing pressure but it also predicts that the rate of increase will decrease with increasing pressure, which is contrary to our expectations. We have suggested an empirical exponential equation for wave speed which overcomes both of these problems and has the added benefit of fitting the detailed experimental measurements wave speed as a function of pressure in the canine aorta made by Anliker and his colleagues.\cite{Anliker1968}\cite{Anliker1972}\cite{Histand1973} This equation should be easy to implement in 1-D numerical studies of arterial hemodynamics and it will be interesting to see how it affects their predictions.
 
 Probably the most common assumption about arterial wave speed is that it is constant as the pressure changes. This is seldom stated but is implicit in virtually every method that has been suggested for determining the local wave speed from measurements of pressure (or area, or diameter) and velocity (or volume flow rate). Impedance analysis of arterial hemodynamics does not make this assumption and generally assumes that the waves in arteries are dispersive, i.e. the wave speed depends upon the frequency of Fourier component $c = c(\omega)$. Because calculation of the impedance, the ratio of pressure to volume flow rate, depends on linearity it is not possible to have a pressure dependent wave speed. On the other hand, the method of characteristics solution of the 1-D conservation equations, in the absence of dissipation, predicts that all waves will propagate at the local wave speed but allows for the possibility that $c = c(P)$.
 
 One of the main uses of $c$ in wave intensity analysis is in the separation of the measured $P$ and $U$ waveforms into their forward and backward components $P_\pm$ and $U_\pm$. This is done using the water-hammer equations (\ref{eq:water_hammer}). The separation is valid if $c=c(P)$ (always assuming that the forward and backward waves are additive, the acoustic limit). For additive waves (\ref{eq:water_hammer}) implies that
 \[
  dP_\pm = \half\left(dP \pm \rho c(P)dU\right)
 \]
 To find $P_\pm$ waveforms, it is necessary to integrate (sum) $dP_\pm$ from $t=0$ to $t$, taking into account the pressure dependency of $c$. This, of course, can be done but complicates the analysis considerably. This can also be seen in the method of characteristics solution where the directions of the characteristics, $\frac{dx}{dt} = U(t) \pm c(P(t))$, are not straight lines in the $x-t$ plane but curves that depend upon the current values of $U$ and $P$. The effects of these non-linearities, convection and pressure dependent wave speed, on arterial hemodynamics have received little attention (although both effects are incorporated into 1-D computer models of the circulation).
 
 \subsection{limitations of the study}
 
 This work is not intended as a review of arterial tube laws and equations for wave speed; it is intended as a critique of the most commonly used tube laws and wave speed equations. Many other relationships have been put forward, but are not commonly used. The discussion of the experimental validity of the various laws and equations are limited to a single example of results for wave speed as a function of pressure the the dog aorta by Aniker \textit{et al.}\cite{Histand1973}. This is mainly because, as far as we know, equivalent data are not available for humans (or other animals). This is a serious deficiency of this work which could be rectified by further experimental studies. Similarly the empirical exponential equation for wave speed as a function of pressure is only compared to the same experimental data. Its validity in general is only speculative. Our observations that this equation fits with our expectations about wave speed increasing with arterial pressure and the rate of increase increasing with pressure only demonstrate consistency not validity.

\begin{appendices}

\section{The speed of sound in air: a cautionary tale}

Isaac Newton derived an equation for the speed of sound in air in 1668 (Proposition XLVIII, Theorem XXXVIII).\cite{Newton1687}
 \[
  a^2 = \frac{P}{\rho}
 \]
His theory was based on the newly published Boyle's Law relating pressure and density and he gave the value 295 m/s (in modern SI units). In the second edition of \textit{Principia} he gave the same formula but an increased value for the speed of sound using more recent values of density as a function of pressure. In the century following a number of variants on his formula for the speed of sound were given by Euler (1727), Lagrange (1759), Biot (1802), Poisson (1808) and a host of others. All of them assumed that acoustic waves are isothermal and all of the formulas gave results that were smaller than the observed speed of sound.

Laplace (1823) first proposed that sound waves are adiabatic rather than isothermal and derived the modern formula 
\[
 a^2 = \frac{\gamma P}{\rho}
\]
where $\gamma$ is the ratio of the specific heat of air at constant pressure and constant volume. This theory was hotly debated for more than 50 years and was not generally accepted for more than 200 years after Newton.

The speed of sound was measured to within 2\% of the currently accepted value, 343 m/s (dry air at 20$^o$ C) by the Accademia del Cimento in Florence in 1660.\cite{Truesdell1979}

It would be tragic if the differences between our theoretical and experimental understanding of wave speeds in arteries takes as long to resolve.

\section{A phenomenological derivation of the equation for wave speed}

Assume that an infinitesimal wave front is propagating in an infinite, uniform tube filled with an inviscid incompressible fluid. Assume that the pressure, velocity and the area before the arrival of the wave are $P_0$, $U_0$ and $A_0$. Assume that the wave propagates with speed $\pm c$ (the sign depends on the direction of travel of the wave) and is convected by the uniform velocity in the tube. Also assume the the changes in pressure, velocity and area as the wave passes are $\Delta P$, $\Delta U$ and $\Delta A$. In coordinates attached to the tube, this problem is non-steady and difficult to analyse. However, if we define coordinates moving with the wave front the problem becomes steady and can be analysed easily using the conservation of mass and energy.

In the wave coordinates the pressure and area are not changed but the velocity becomes $U = \mp c$ upstream of the wave and $U = \mp c + \Delta U$ downstream of the wave. By the conservation of mass for an incompressible fluid the volume flow rate must be constant on both sides of the wave.
\[
 UA = \mp cA_0 = (\mp c+\Delta U)(A_0 + \Delta A)
\]
Neglecting terms of $\mathcal{O}(\Delta^2)$, conservation of mass requires
\[
 A_0\Delta U \mp c\Delta A = 0
\]
By the conservation of energy, remembering that pressure has the units of energy per unit volume, the total pressure $P +\frac{1}{2}\rho U^2$ must also be constant on both sides of the wave. If the fluid is incompressible
\[
 P + \half \rho U^2 = P_0 + \half \rho (\mp c)^2 = P_0 + \Delta P + \half \rho ((\mp c+\Delta U)^2
\]
Again neglecting terms of $\mathcal{O}(\Delta^2)$ conservation of energy requires
\[
 \Delta P \mp \rho c \Delta U = 0
\]
This equation gives the relationship between the change in pressure and the change in velocity across the wave front, generally known as the water-hammer equations
\[
 \Delta P_\pm = \pm \rho c \Delta U_\pm
\]
where the subscript denotes the direction of travel of the wave. If we use the water-hammer equations to eliminate $\Delta U$ from the equation arising from the conservation of mass, we obtain an equation for the wave speed
\[
 c^2 = \frac{A_0 dP}{\rho \Delta A}
\]
which is the equation for the wave speed derived from the solution of the 1-D partial differential conservation equations by the method of characteristics (\ref{eq:c_def}). This remarkably simple analysis produces both the equation for $c$ as a function of $P$ and $A$ and the water-hammer equations relating $\Delta P$ and $\Delta A$ across the wave front.

\end{appendices}

\end{document}